\documentclass{article}
\usepackage{spconf,amsmath,graphicx}
\usepackage{url}

\title{SLAP: A Split Latency Adaptive VLIW Pipeline Architecture which enables on-the-fly variable SIMD vector-length} 	
\name {Ashish Shrivastava$^{\star}$$^{\dagger}$, Alan Gatherer$^{\star}$$^{\ddagger}$, Tong Sun$^{\|}$, Sushma Wokhlu$^{\|}$, Alex Chandra$^{\|}$  \thanks{${\|}$Authors performed the work while at Futurewei Technologies Inc}}
\address{$^{\star}$Wireless Access Lab, Futurewei Technologies Inc\\ 
 $^{\dagger}$Senior Member, IEEE \qquad $^{\ddagger}$Fellow, IEEE
}
 
\begin{document}

\begin{titlepage}
\raggedright
© 20XX IEEE. Personal use of this material is permitted. Permission from IEEE must be obtained for all other uses, in any current or future media, including reprinting/republishing this material for advertising or promotional purposes, creating new collective works, for resale or redistribution to servers or lists, or reuse of any copyrighted component of this work in other works.
\end{titlepage}
\pagebreak

%
\maketitle
\begin{abstract}
Over the last decade the relative latency of access to shared memory by multicore increased as wire resistance dominated latency and low wire density layout pushed multi-port memories farther away from their ports. Various techniques were deployed to improve average memory access latencies, such as speculative pre-fetching and branch-prediction, often leading to high variance in execution time which is unacceptable in real-time systems. Smart DMAs can be used to directly copy data into a layer-1 SRAM, but with overhead. The VLIW architecture, the de-facto signal-processing engine, suffers badly from a breakdown in lock-step execution of scalar and vector instructions. We describe the Split Latency Adaptive Pipeline (SLAP) VLIW architecture, a cache performance improvement technology that requires zero change to object code, while removing smart DMAs and their overhead. SLAP builds on the Decoupled Access and Execute concept by 1) breaking lock-step execution of functional units, 2) enabling variable vector length for variable data-level parallelism, and 3) adding a novel triangular-load mechanism. We discuss the SLAP architecture and demonstrate the performance beneﬁts on real traces from a wireless baseband-system (where even the most compute intensive functions suffer from an Amdahl’s law limitation due to a mixture of scalar and vector processing). 
\end{abstract}

\section{Introduction \& Prior Work}
In wireless Baseband SoC architecture \cite{Marvell/Octeon_CNF73xx}\cite{Marvell/Octeon_CNF95xx}\cite{Ericsson/ASIC&SoC} programmable compute engines are widely used in physical layer implementation and are split into three categories 1) control/scalar dominant (integer arithmetic) requiring a CPU 2) control/scalar along with floating-point signal processing with limited data-level parallelism requiring narrow vector floating-point units, and 3) control/scalar processing along with heavy floating-point signal processing (with higher data-level parallelism) requiring wider vector floating-point units. VLIW has become the de-facto processor technology for narrow SIMD and wider SIMD processors but introduces a fundamental limitation if implemented traditionally (i.e. all functional units in lock-step execution) when the algorithm calls for a mixture of scalar functional units with integer arithmetic (and therefore smaller pipeline stages) and vector/SIMD functional units with floating-point capabilities (and higher execute pipeline stages) with strict dependency between scalar functionality and vector/SIMD functionality. For example, scalar functional units calculate addresses for the floating-point data that will be processed by vector functional units. For wireless baseband, relaxing the VLIW’s lock-step coupling between the scalar and vector parts reduces cycle count and makes the architecture scalable to different vector lengths. To support all the required 5G physical layer numerology/use-cases, the memory infrastructure is hierarchical, non-uniform, non-coherent and distributed with on-chip multi-layer caches \& SRAMs as well as off-chip DRAMs \cite{ti8}\cite{fsc9}\cite{ceva10}\cite{ceva11}.
\begin{figure}
	\centering
	\begin{minipage}{.8\columnwidth}
		\includegraphics[width=1\linewidth]{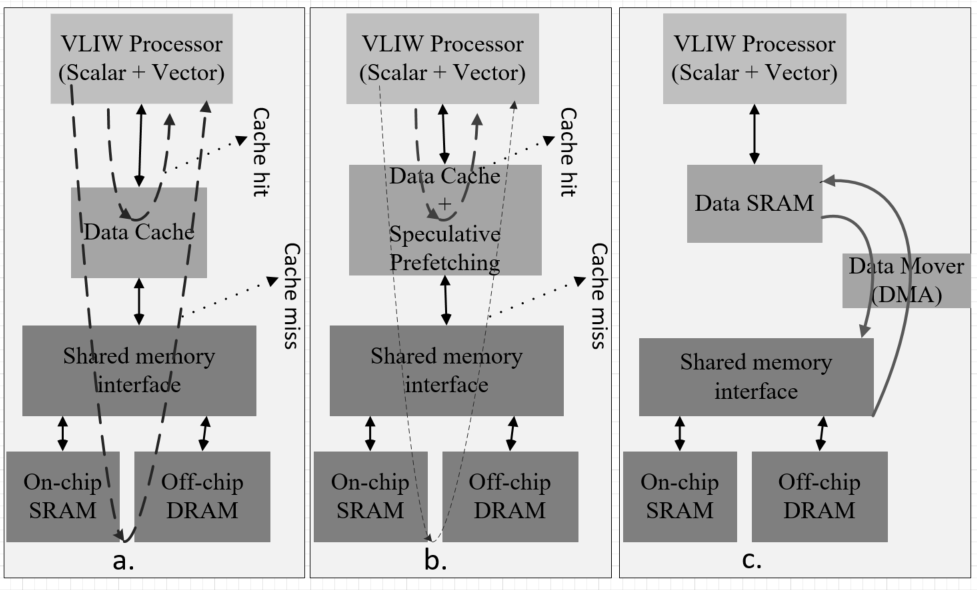}
	\end{minipage}
	\begin{minipage}{.6\columnwidth}
		\includegraphics[width=1\linewidth]{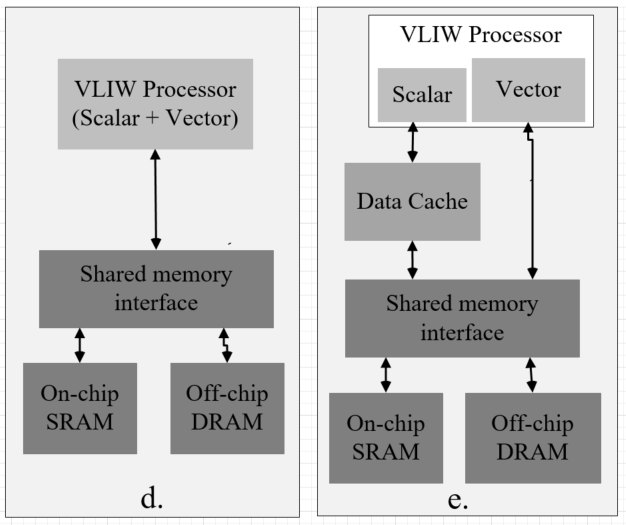}
	\end{minipage}
	\caption{Processor with hierarchical memory architecture}
	\label{fig:0}
\end{figure}
In Fig. \ref{fig:0}.c we see that a non data cached system \cite{ceva10} allows the VLIW machine to run in an efficient and predictable manner but with the overhead of programmed data movement operations in the DMA. This impacts power and latency unless the data movement can be pipelined perfectly with the computation; a task that proves to be very difficult due to the runtime variety and complexity of a modern modem. For this reason, all the major suppliers of baseband VLIW DSPs \cite{ti8}\cite{ceva11} provide cache support to it, as shown in Fig. \ref{fig:0}.a. The code becomes more portable and robust but performance degrades due to misses on data accesses because data accesses dimensionality is flexible at runtime, and there are usually some “rogue” parameters that need to be accessed within the loop that may come from tables that are too big to be stored in L1 cache or perhaps even in L2. It is therefore difficult to hand optimize the loops or successfully use speculative pre-fetching techniques as shown in \ref{fig:0}.b. Much of the baseband physical layer code does not show “temporal locality of reference” and the “spatial locality of reference” gets polluted at runtime, leading to very minimal benefit from complex cache logic.
To address these concerns \cite{hwacha13} combines ideas from DAE \cite{DAE4} and decoupled vector architectures \cite{DVA5} to improve data access for large SIMD as shown in Fig. \ref{fig:0}.d where prefetching logic on a single processor accesses scalar and vector data. \cite{PREDICT_LDST} identifies load instructions that impact the cache misses allowing selective prefetching but this assumes static and predictable code traversal. \cite{qual12} follows Fig. \ref{fig:0}.e with scalar threads accessing memory via L1 caches and vector threads bypassing L1 Cache but this requires vector prefetch optimization in the code. SLAP follows Fig. \ref{fig:0}.e, but differs from \cite{qual12} in that we allow multiple lanes to access the memory in an un-synchronized manner using elastic queues to manage correct operation. 
SLAP’s split pipeline architecture uses the DAE principle but does so in a code transparent manner whereas \cite{DAE4} is a accelerator implemented as a vector extension along with a separate control processor in which the scalar and vector units access memory directly whereas the control processor has data cache so the decoupling of scalar and vector processing happens at compile time with two sets of programs, one for control and another for vector execution. Similarly, \cite{PIPE} implements the DAE \cite{DAE4} with two sets of processors and two sets of programs producing the same scheduling problems that appear when there is a separate L1 load engine. SLAP’s architecture on the other hand has a single program optimized at compile time (like traditional VLIW) and the vector pipeline splits away at runtime in the hardware (including vector load/store instructions) allowing for better code portability and object code compatibility. \cite{DYNAMIC_VLIW} proposes split-issue in conjunction with delay-buffer and reservation stations to support dynamic scheduling mechanism and the purpose is similar to SLAP. But it requires costly (in terms of power and area) reservation stations to support different access policies and out-of-order pipeline execution. SLAP split pipeline architecture is in-order and shows significant performance improvement with no additional power and area overhead. 

\section{Architecture Overview}
\begin{figure}
	\centering 
	\begin{minipage}{.7\columnwidth}
		\includegraphics[width=\linewidth]{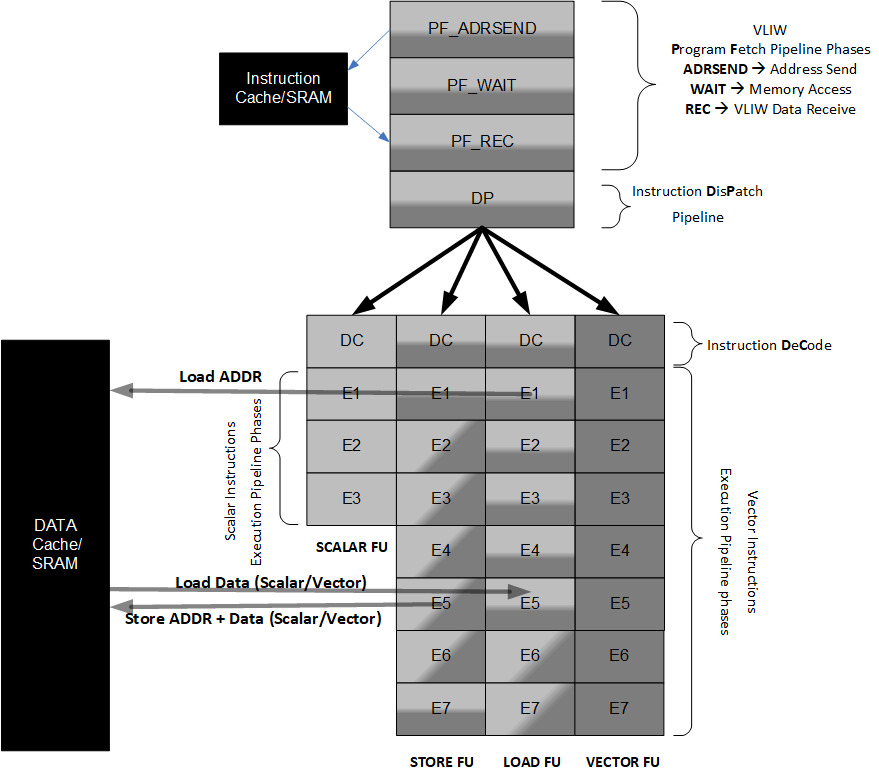}
	\end{minipage}
	\caption{Typical VLIW processor pipeline stages}
	\label{fig:1}
\end{figure}

We compare our SLAP VLIW processor to the in house VLIW DSP from which it is derived by adding SLAP features without changing the pipeline. This allows common object code for fair comparison. Also, it is important to minimize changes to object code as this places a heavy burden on code maintenance. Fig. \ref{fig:1} shows typical VLIW pipeline stages with a 3 phase Fetch stage (addressing, memory access, and data), instruction dispatch (DS), instruction decode (DC), and an Execute pipeline with different stages for scalar processing and SIMD floating-point processing.
Note that the pipeline requires functional units to execute in lock-step. The SIMD floating point pipeline executes matrix and vector operations and differs dramatically from the scalar integer arithmetic pipeline which typically executes address calculations for memory operations and software-pipeline loop counts. There is relatively little data exchange between these pipelines which traditionally operate in lock-step. So when a scalar operation is stalled (due to memory read operations) it stalls the vector SIMD functional units, and vice versa.
\subsection{SLAP based VLIW processor Pipeline overview}
\begin{figure}
	\centering 
	\begin{minipage}{.8\columnwidth}
		\includegraphics[width=\linewidth]{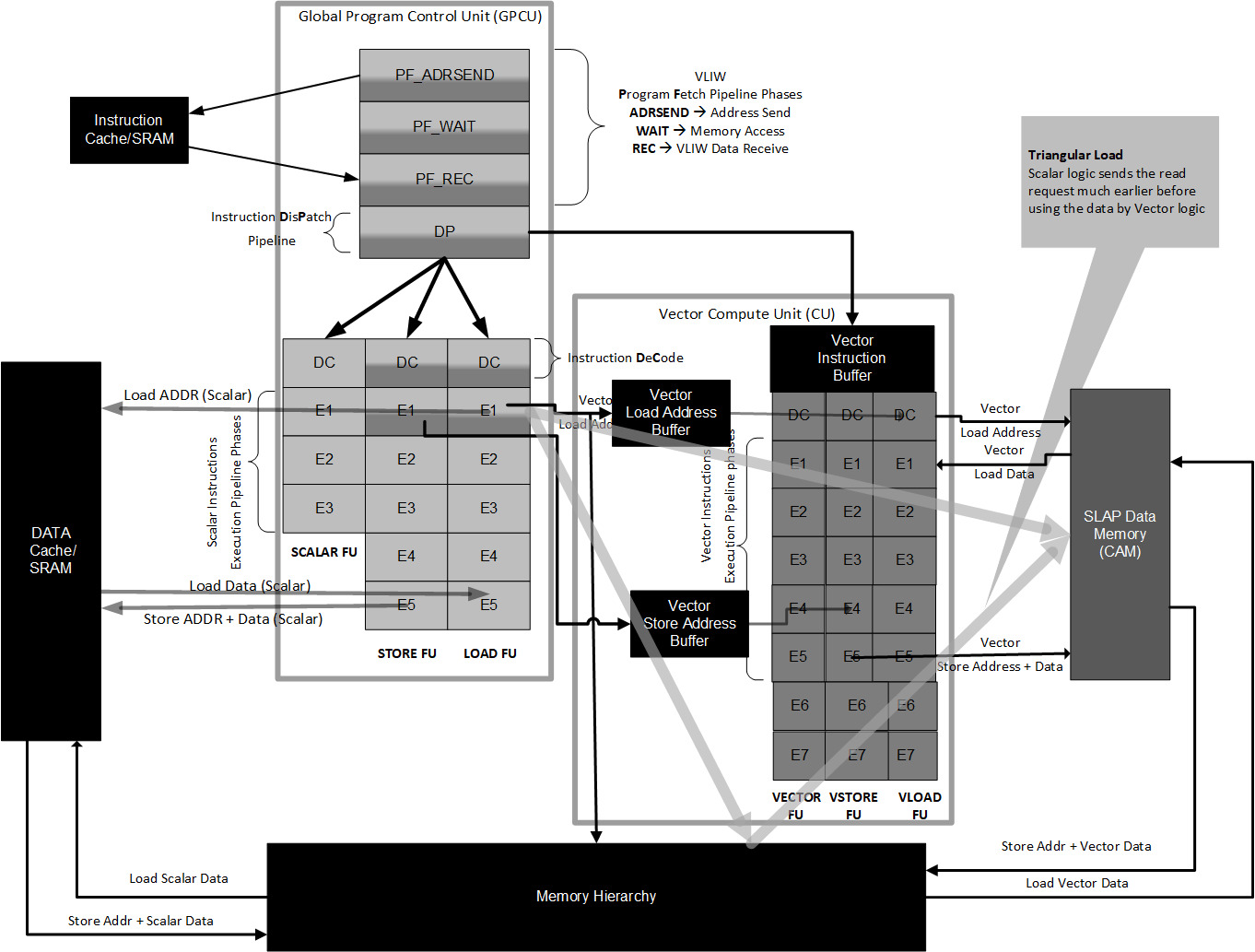}
	\end{minipage}
	\caption{SLAP based VLIW Architecture}
	\label{fig:2}
\end{figure}

For wireless baseband, memory communication latency stalls significantly impact performance by stalling scalar and vector pipelines. Techniques to improve average memory stalls, such as speculative/selective prefetching and branch prediction are very complex and often under-utilized, while floating point operations have minimal data exchange with scalar operations. These observations point to the potential benefit of relaxing timing constraints between scalar a\label{key}nd vector pipelines by allowing variable-lag (but controlled) execution, to improve throughput by mitigating the impact of costly memory communication delays. This is achieved in SLAP by “dispatching” (during Instruction dispatch pipeline stage) the vector arithmetic operations into a Vector Instruction buffer (with appropriate control logic to throttle the vector instruction flow). This is shown in Fig. \ref{fig:2}, for the scalar pipeline, called the GPCU (Global Program Control Unit) and a single vector pipeline, called a CU (compute unit). Other CUs are easily added. attaching to the GPCU and memory in the same way, with each CU having an independent port to memory. SLAP allows the variable offset/lag (up to the size of the Vector Instruction Buffer depth) execution between GPCU and CU. The Vector instruction buffers are implemented as low power FIFOs as shown in section \ref{perf_study}. 
Typically a single Load/Store functional unit performs scalar and vector loads/stores and address updates along with register forwarding to scalar and vector register files. In SLAP, the Load/Store pipeline is split, one for scalar arithmetic (along with scalar-register forwarding logic), and another for vector arithmetic floating-point (and vector-register forwarding). The address generation for the CU (vector floating-point operations) are part of the scalar functionalities and therefore all the address calculations for all CU loads/stores are done in the GPCU and the addresses are transferred to CU via the vector Load Address buffer and the vector store address Buffer. Vector load/store instructions are split into three categories 1) GPCU managing the address for vector-SIMD loads/stores as well as implementing a “Triangular load” mechanism \cite{patent7} for Vector loads, 2) the CU interprets those vector SIMD load instructions as reading from the SLAP Data memory, and 3) the CU interprets vector stores instructions as vector stores with address popped from the write address buffer.
\subsection{SLAP triangular load overview}
SLAP's triangular load is a novel way of implementing a DAE \cite{DAE4}. In a typical VLIW processor, a load functional unit will issue the read memory instructions and received the data into the shared register files for consumption. If the data is not available within the pipeline stages of the read instruction, the whole processor stalls. Triangular load allows the GPCU, running ahead of the CU, to issue reads and the data to return to the CU Data Memory CAM (Content-Addressable Memory \cite{cam}), as shown in Fig. \ref{fig:2}.  
\subsection{SLAP based Variable SIMD architecture overview}
One of the key features of VLIW architecture, compiler controlled static scheduling, is actually a curse in disguise. Basestation applications require different categories of computations (scalar-only, narrow-SIMD and wide-SIMD). This leads to multiple compiler/debug toolchains, one for each type of VLIW but derived from the base-line VLIW processor. Since the software uses intrisics\cite{VLIW_BOOK} to specify the SIMD instructions and unique datatypes\cite{dsp_datatypes}, lots of software needs to be rewritten and verified when migrating from one VLIW processor to another. SLAP removes this issue by enabling variable vector/SIMD for variable data-level parallelism with a single compiler/debug tool-chains.
\begin{figure}
	\centering 
	\begin{minipage}{.4\columnwidth}
		\includegraphics[width=\linewidth]{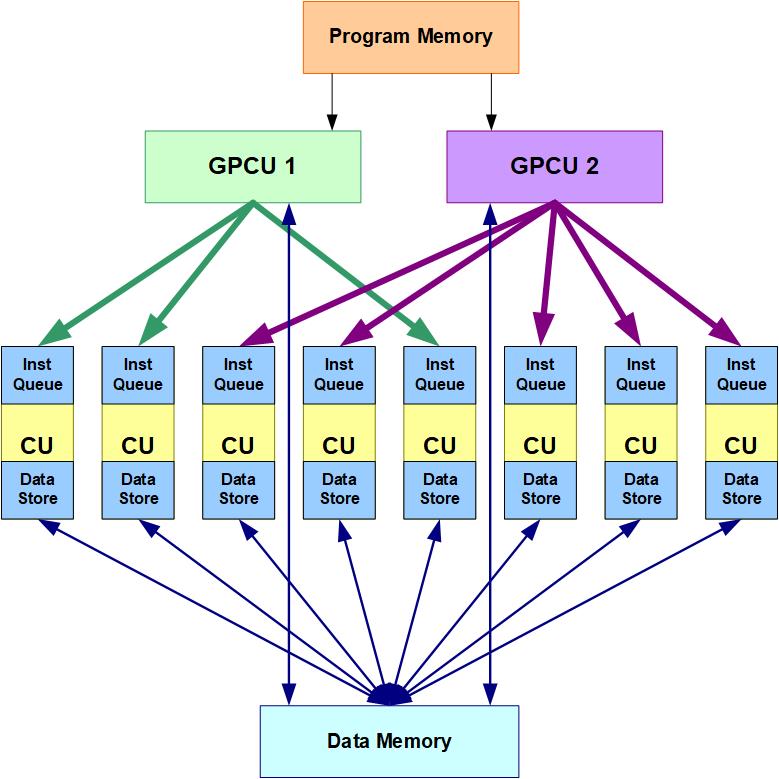}
	\end{minipage}
	\caption{SLAP based Variable SIMD Vector architecture}
	\label{fig:7}
\end{figure}
SLAP allows a variable SIMD/vector VLIW architecture via dynamic resource sharing as shown in Fig. \ref{fig:7} by allowing  the association of GPCUs to CUs dynamically, enabling variable vector length on the fly. Fig. \ref{fig:7} shows two GPCUs sharing 8 CUs with a dynamic configuration where GPCU1 is controlling 3 CUs and GPCU2 is controlling 5 CUs. If each CU supports SIMD4 then GPCU1 processes SIMD12 and GPCU2 processes SIMD20. The number of CUs assigned depends on the number of distinct data sets to process.  The interface between the GPCUs and the CUs is a FIFO of instructions at each CU. Instructions are pushed onto the FIFO simultaneously by the GPCU, but each CU may pop instructions at their own speed as they are ready to execute them.  If an instruction queue fills, the GPCU stalls and doesn’t dispatch any more instructions until space is available in all of the CU instruction queues.  If a CU instruction queue empties, the CU stalls and waits for additional instructions from the GPCU. Both of these cases may happen during normal operations.
\section{Performance Study}{\label{perf_study}
We developed cycle-accurate models for the DSP processor, caches, and memory hierarchy and created a multi-processor SoC model for regular and SLAP DSPs. For performance analysis, we collected execution traces from the in-house VLIW DSP cores on our Basestation SoCs running Physical Uplink Shared Channel (PUSCH) receive chain and then ran these traces on our SoC model. Fig. \ref{PUSCH_FIG} shows processing blocks in PUSCH receive chain in green (see \cite{3gpp.36.211} for more details). The DSP clusters process the estimation blocks above the main processing flow such as reference extraction, channel estimation, CQI etc. and are also in charge of control and management, and the generations of parameters for the receiver chain. DSP clusters will be involved in other processing, but PUSCH is the most compute intensive and the most appropriate to benchmark DSP performance. 
\begin{figure}
	\includegraphics[width=\linewidth]{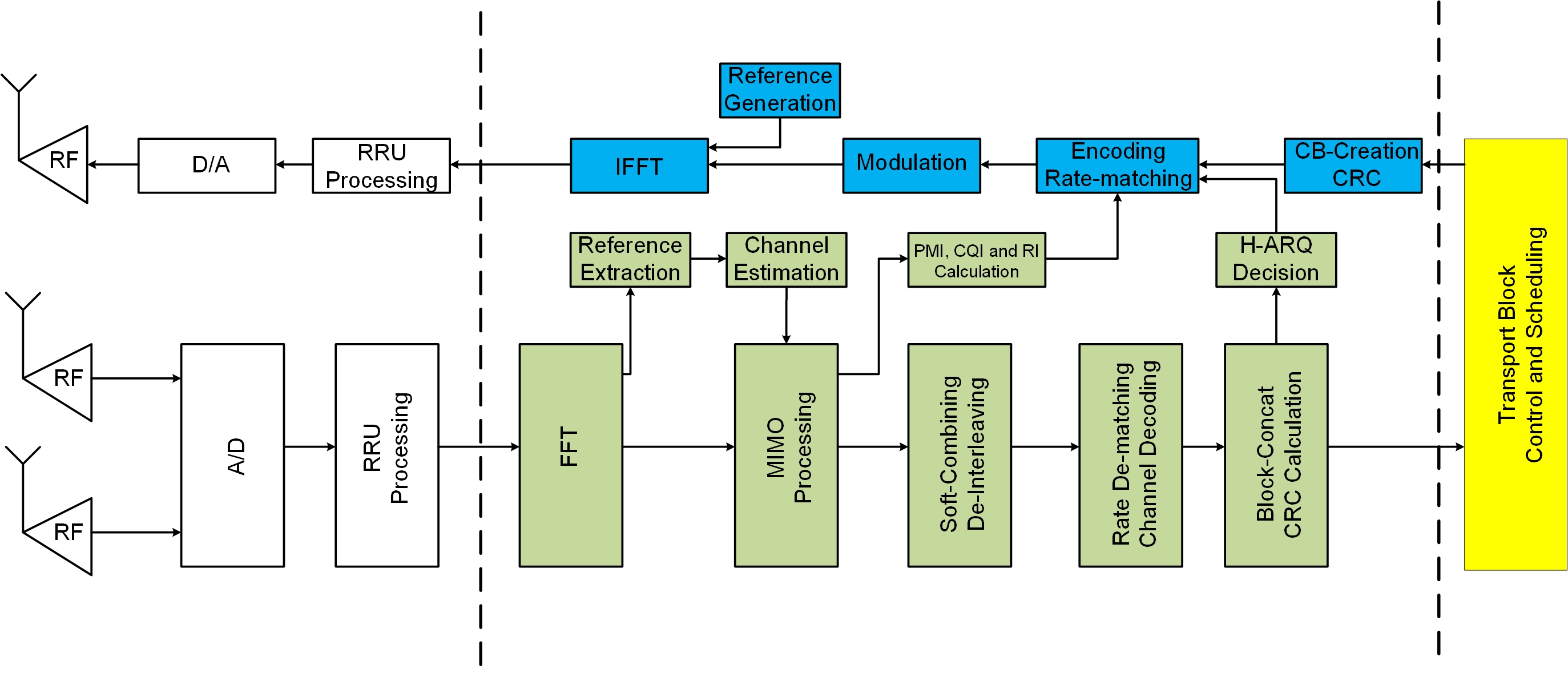}
	\caption{Wireless Baseband PUSCH processing blocks }
	\label{PUSCH_FIG}
\end{figure}
Traces were collected from our product for different regions (based on scalar/vector operations distribution) and stitched together as shown by the regions of performance in Fig. \ref{fig:5}. For example, Region 1, 3, and 5 have higher scalar operations than vector, whereas region 2 and 4 shows mixed scalar and vector operations.
For this "combo trace" we benchmarked SLAP with various CU instruction FIFO sizes and different GPCU data caches sizes as shown in Fig. \ref{fig:4} we tabulate the \% increase in runtime compared to an in-house DSP with flat-memory (so with no cache misses or memory accesses stalls). Adding 32KB data cache (with all speculative/selective prefetching mechanism) to the in house DSP results in about 33.63\% of overhead and SLAP with FIFO size of 24 or 32 combined with GPCU data cache size of 8KB, 16KB or 32KB reduces this overhead with a sweet spot at a FIFO size of 32 and the opportunity to reduce the cache to 16KB without much degradation, though even an 8KB cache produces benefit.
\begin{figure}
	\includegraphics[width=\linewidth]{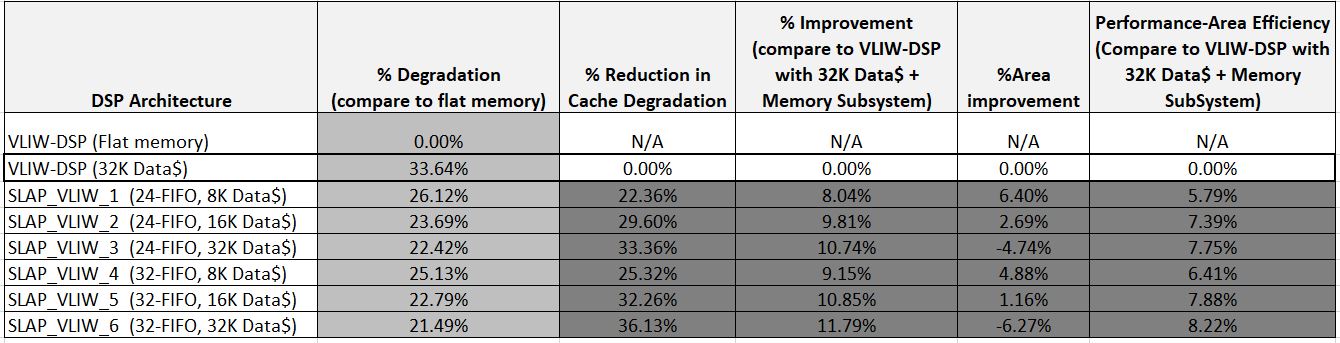}
	\caption{SLAP-VLIW performance/area efficiency for PUSCH}
	\label{fig:4}
\end{figure}
Fig. \ref{fig:4}, shows the performance-area efficiency of PUSCH improves in the range of 5.7\% - 8.22\%. The test-vectors used are the traces captured on the in-house DSP. Recompiling for SLAP gives more improvements but we do not have the space to present them here.
Fig. \ref{fig:5} shows how the ratio of scalar to vector instructions varies across the combo trace and one would expect that the performance benefit would be strongly dependent on this ratio. In \ref{perf_data} we summarize the benefit for SLAP FIFO depth of 24 and 8KB data cache, versus in-house DSP with 32KB data cache, for different regions of the combo trace.
\begin{figure}	
	\begin{minipage}{.7\columnwidth}	
		\centering
		\includegraphics[width=\linewidth]{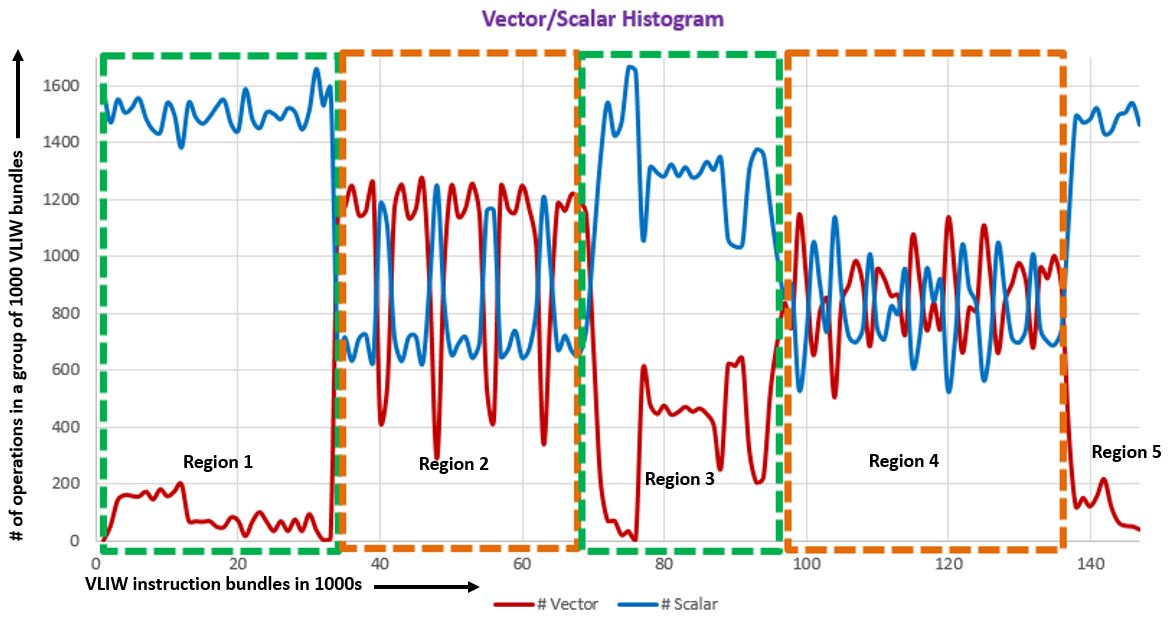}
		\caption{PUSCH Scalar/Vector regions}
		\label{fig:5}		
	\end{minipage}
	\begin{minipage}{.5\columnwidth}
		\centering
		\includegraphics[width=\linewidth]{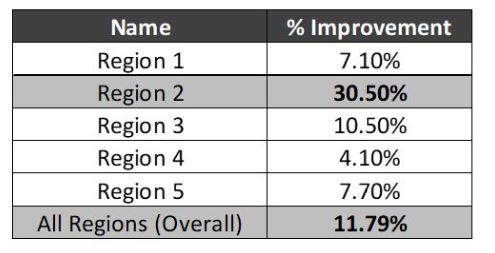}
		\caption{SLAP improvements}
		\label{perf_data}		
	\end{minipage}
\end{figure}
Region 1, 3 and 5 shows heavy scalar processing but the gain of SLAP is still in the range of 7.10\% – 10.50\%. This shows that SLAP is also benefiting scalar processing by not polluting the data cache with vector processing. Region 2 and 4 evenly mix scalar and vector processing but the improvement ratio is 30.6\%. 4.10\% respectively. Clearly SLAP improvement depends on the patterns of scalar/vector processing as much as the simple ratio. But in all cases the benefit is significant.

Though the area of SLAP decreases this is due to a reduction in memory, which is generally cool, and the addition of FIFO logic. One might suspect that the FIFO logic will run much hotter than the memory removed, increasing the overall power. We performed a careful study of the power dissipation of the FIFOs compared to register files of the equivalent size and found that FIFO implementation is over an order of magnitude more power efficient than a similarly sized multi-ported register file. This is because of the dramatic reduction in number of ports and wire length in the memory. In SLAP though we are using long FIFOs we can still save power because we reduce cache size.

\section{Conclusions and Future Work}
We have shown that an extension of a DAE structure can provide significant, double digit percentage performance benefit to a VLIW DSP in a basestation application while potentially decreasing power and certainly decreasing area. This is possible without recompilation of the original DSP code and can also extend to dynamic sharing of pools of SIMD units. There are several areas for future study, including the development of a scheduling strategy to take advantage of the dynamic allocation of SIMD to improve system level scheduling goals.

\vfill\pagebreak
\label{sec:refs}
\bibliographystyle{IEEEbib}
\bibliography{strings,refs,slap_refs}

\end{document}